\documentclass[preprint,prd,tightenlines,nofootinbib,eqsecnum]{revtex4-1}

\usepackage{amsmath}
\usepackage{amsfonts}
\usepackage{amssymb}
\usepackage{bm}
\usepackage{color}
\usepackage[colorlinks,pdfborder={000},plainpages=false]{hyperref}

\definecolor{darkblue}{rgb}{0,0,0.3}
\hypersetup{citecolor=darkblue}
\hypersetup{linkcolor=darkblue}

\newcommand{\beq}{\begin{equation}}
\newcommand{\eeq}{\end{equation}}
\newcommand{\ud}{\mathrm{d}}
\newcommand{\calE}{\mathcal{E}}
\newcommand{\calM}{\mathcal{M}}
\newcommand{\calO}{\mathcal{O}}
\newcommand{\bo}[1]{\mathbf{#1}}
\newcommand{\av}[1]{\langle #1 \rangle}

\allowdisplaybreaks

\begin{document}

\title{First Law of Mechanics for Compact Binaries on Eccentric Orbits}

\author{Alexandre Le Tiec}

\affiliation{Laboratoire Univers et Th\'eories, Observatoire de Paris, CNRS, Universit\'e Paris Diderot, 92190 Meudon, France}

\date{\today}

\begin{abstract}
Using the canonical Arnowitt-Deser-Misner Hamiltonian formalism, a ``first law of mechanics'' is established for binary systems of point masses moving along generic stable bound (eccentric) orbits. This relationship is checked to hold within the post-Newtonian approximation to general relativity, up to third (3PN) order. Several applications are discussed, including the use of gravitational self-force results to inform post-Newtonian theory and the effective one-body model for eccentric-orbit compact binaries.
\end{abstract}


\maketitle

\section{Introduction}

The direct observation of gravitational waves would have a tremendous impact on physics, astrophysics and cosmology \cite{SaSc.09}. Binary systems composed of compact objects, namely black holes and neutrons stars, are highly promising sources of gravitational waves \cite{BuSa.15}. The orbital dynamics of these systems can be investigated by means of various approximation methods in general relativity, such as post-Newtonian (PN) theory \cite{FuIt.07,Sc.11,FoSt.14,Bl.14}, the perturbative gravitational self-force (GSF) approach \cite{Ba.09,Po.al.11,Th.11,Ba.14} and the effective one-body (EOB) framework \cite{BuDa.99,BuDa.00,Da.14}. These approximation methods are complementary to fully nonlinear numerical simulations of the late inspiral and merger of compact binaries. Recent years have seen a spur of activity at the multiple interfaces of these various techniques. Such cross-cultural studies have contributed to improving our knowledge of the two-body dynamics and
wave emission \cite{Le2.14}.

Some of this progress stems from the ``first law of binary mechanics'' of Le Tiec \textit{et al.} \cite{Le.al.12} (hereafter Paper I), a variational formula that relates small changes in the bodies' masses to those of the total mass-energy and angular momentum of a binary system moving along a circular orbit. This variational relationship is a particular case, valid when one assumes the existence of a global helical Killing field and a point-particle model for the compact objects,\footnote{Mathematically, the approximation of an exactly closed circular orbit translates into the existence of a helical Killing vector field, along the orbits of which the spacetime geometry is invariant.} of the generalized first law of mechanics of Friedman \textit{et al.~}\cite{Fr.al.02}. More recently, Blanchet \textit{et al.} \cite{Bl.al.13} have extended to spinning point particles the first law of Paper I, for spins aligned or anti-aligned with the orbital angular momentum.

The first law of Paper I was successfully applied to (i) the determination of the numerical values of some previously unknown PN coefficients (at 4PN, 5PN and 6PN orders) that enter the expressions for the binding energy and angular momentum as functions of the circular-orbit frequency \cite{Le.al.12}, (ii) the derivation of the exact expressions for the binding energy and orbital angular momentum at linear order in the mass ratio \cite{Le.al2.12}, and (iii) the computation of the frequency shift of the Schwarzschild innermost stable circular orbit (ISCO) frequency induced by the conservative piece of the GSF \cite{Le.al2.12,Ak.al.12}. Moreover, it was used in Refs.~\cite{Ba.al.12,Ak.al.12} to compute one of the potentials that enters the EOB effective metric, exactly, at linear order in the mass ratio.

The first laws of Refs.~\cite{Le.al.12,Fr.al.02,Bl.al.13} were established for compact binaries moving along circular orbits. In this paper, we shall extend the first law of Paper I to binary systems of nonspinning compact objects moving along \textit{generic} stable bound (eccentric) orbits. There are multiple reasons to do so. First, although most stellar-mass compact binaries would have completely circularized by the time they enter the observable frequency band of ground-based detectors such as Advanced LIGO, Advanced Virgo and KAGRA, there are scenarii where eccentricity effects could become observable and would give access to much interesting physics \cite{We.03,OL.al.09,Th2.11,KoLe.12,AnPe.12,Sa.al.14}. Second, eccentric inspirals with extreme mass ratios are promising sources for a future mHz-band gravitational-wave antenna in space such as the proposed eLISA mission \cite{Am.al.13,GaPo.13,Am.al2.13,Am.al.15,Ba.al.15}. \!Third, eccentric orbits give access to new degrees of freedom in the EOB model \cite{Da.10}.

The remainder of this paper is organized as follows. The first law of binary mechanics for eccentric-orbit binaries is derived in Sec.~\ref{sec:derivation}, while some of its consequences are explored in Sec.~\ref{sec:consequences}. This relationship is checked to hold true, in the context of the PN approximation, up to 3PN order, in Sec.~\ref{sec:verification}. Finally, Sec.~\ref{sec:applications} is devoted to applications, such as the use of GSF results to inform PN theory and the EOB model for eccentric-orbit compact binaries. Throughout this paper we use ``geometrized units'' where $G = c = 1$.

\section{Derivations of the first law}\label{sec:derivation}

In this section, building upon the canonical Arnowitt-Deser-Misner (ADM) Hamiltonian formalism (Sec.~\ref{subsec:ADM}), we derive a first law of mechanics for binary systems of compact objects moving along eccentric orbits. The first proof relies on an orbital averaging (Sec.~\ref{subsec:average}), while the second proof makes use of action-angle type variables (Sec.~\ref{subsec:q_J}).

\subsection{Canonical Hamiltonian of point-particle binaries}\label{subsec:ADM}

Our starting point is the application of the ADM canonical formulation of general relativity \cite{Ar.al.62} to a binary system of nonspinning compact objects, modelled as point particles with constant masses $m_a$ (with $a=1,2$), as recently reviewed in Refs~\cite{Sc.11,Sc.14}. The orbital dynamics of this binary system is assumed to derive from an autonomous Hamiltonian $H(\bo{x}_a,\bo{p}_a;m_a)$, where $\bo{x}_a(t)$ and $\bo{p}_a(t)$ are the canonical positions and momenta of the two particles, defined in some fixed gauge. Such a Hamiltonian has been computed, up to 4PN order,\footnote{Starting at 4PN order, the binary Hamiltonian involves some nonlocal-in-time contribution, such that $H$ becomes a functional of the phase-space variables $\bo{x}_a(t)$ and $\bo{p}_a(t)$.} using the ADM transverse-traceless (TT) gauge \cite{JaSc.12,JaSc.13,Da.al.14}. The phase-space coordinates $\bo{x}_a(t)$ and $\bo{p}_a(t)$ obey Hamilton's equations
\beq\label{Hamilton}
	\dot{\bo{x}}_a = \frac{\partial H}{\partial \bo{p}_a} \, , \quad \dot{\bo{p}}_a = - \frac{\partial H}{\partial \bo{x}_a} \, ,
\eeq
where Euclidean three-vectors are denoted in bold font, and an overdot stands for the derivative $\ud / \ud t$ with respect to coordinate time $t$.

In general relativity, asymptotically flat spacetimes possess ten conserved quantities, given as surface integrals at spatial infinity, that are associated---\textit{via} Noether's theorem---with the continuous symmetries of the Poincar\'e group \cite{ReTe.74,HaReTe}. These quantities are the total energy $H$, total linear momentum $\bo{P}$, total angular momentum $\bo{L}$, and boost vector $\bo{K} = \bo{G} - \bo{P} \, t$, where $\bo{G}$ is the center-of-mass vector. These act as ``generators'' of time translations, spatial translations, spatial rotations, and Lorentz boosts, respectively. For binary point-particle spacetimes, the global Poincar\'e symmetry is realized by the Poincar\'e algebra satisfied by the ten generators $H,\bo{P},\bo{L}$ and $\bo{K}$, viewed as functions on the two-body phase space $(\bo{x}_a,\bo{p}_a)$ \cite{Da.al2.00}. If the coordinate system used to define the canonical Hamiltonian manifestly respects the Euclidean group, such as the usual ADM-TT gauge conditions, then $H(\bo{x}_a,\bo{p}_a;m_a)$ must be translationally and rotationally invariant, which implies \cite{Da.al2.00}
\beq
	\bo{P} = \sum_a \bo{p}_a \, , \quad \bo{L} = \sum_a \bo{x}_a \times \bo{p}_a \, .
\eeq

Thereafter, we shall consider only the \textit{relative} motion of the binary system with respect to the \textit{center-of-mass} frame, which is defined by the condition $\bo{G} = \bo{0}$, implying $\bo{P} = \dot{\bo{G}} = \bo{0}$. This restriction decreases by six the number of degrees of freedom of the dynamical system. Because of translational invariance, the center-of-mass Hamiltonian $H(\bo{r},\bo{p};m_a)$ is a function of the relative position $\bo{r} \equiv \bo{x}_1 - \bo{x}_2$ and the relative momentum $\bo{p} \equiv \bo{p}_1 = - \bo{p}_2$ only. Thus, in the center-of-mass frame, the conserved orbital angular momentum simply reads $\bo{L} = \bo{r} \times \bo{p}$, such that the orbital motion is confined to the coordinate plane orthogonal to $\bo{L} \equiv L \, \hat{\bo{L}}$ and spanned by $\bo{r}(t)$ and $\bo{p}(t)$ at any given time $t$. Introducing polar coordinates in this plane, such that $\bo{r} = (r\cos{\varphi}, r\sin{\varphi},0)$ with $r = \vert \bo{x}_1 - \bo{x}_2 \vert$ the separation and $\varphi$ the orbital phase, we have $L = p_\varphi = \text{const.}$ Therefore, Hamilton's equations \eqref{Hamilton} reduce to

\beq\label{EOM}
	\dot{r} = \frac{\partial H}{\partial p_r} \, , \quad \dot{p}_r = - \frac{\partial H}{\partial r} \, , \quad \dot{\varphi} = \frac{\partial H}{\partial L}  \, , \quad \dot{L} = - \frac{\partial H}{\partial \varphi} = 0 \, ,
\eeq
and the center-of-mass Hamiltonian $H(r,p_r,L;m_a)$ is independent of the cyclic cordinate $\varphi$.

Henceforth, we shall consider only \textit{bound} (and stable) orbits, for which the radial motion is bounded: $r \in [ r_\text{min}, r_\text{max} ]$, where $r_\text{min}$ and $r_\text{max}$ denote the coordinate separation at periastron and apastron, respectively. These two turning points of the radial motion correspond to the two largest (real, positive, and finite) roots of the equation $\dot{p}_r = - \partial H (r,p_r,L;m_a) / \partial r = 0$, evaluated at $p_r = 0$. Since the binary dynamics is conservative, the radial motion of a bound orbit must be periodic in time, such that $r(t+P) = r(t)$ and $p_r(t+P) = p_r(t)$ for all time $t$, where $P$ denotes the coordinate time period of the radial motion.

\subsection{First law from an orbital averaging}\label{subsec:average}

Under infinitesimal changes $\delta r$, $\delta p_r$, $\delta L$ and $\delta m_a$ of the phase-space coordinates and mass parameters, the Hamiltonian undergoes a variation
\beq
	\delta H = \frac{\partial H}{\partial r} \, \delta r + \frac{\partial H}{\partial p_r} \, \delta p_r + \frac{\partial H}{\partial L} \, \delta L + \sum_a \frac{\partial H}{\partial m_a} \, \delta m_a \, ,
\eeq
where to simplify the notation we do not indicate that the partial derivatives with respect to $r$, $p_r$, $L$ and $m_a$ are computed while keeping the other variables fixed.

For two point particles interacting only through (Einsteinian) gravitation, it was shown in Ref.~\cite{Bl.al.13} that the partial derivatives of the canonical ADM Hamiltonian with respect to the particle's masses, holding the phase-space coordinates fixed, are simply given by
\beq\label{z_a}
	\frac{\partial H}{\partial m_a} = \dot{\tau}_a \equiv z_a \, ,
\eeq
where $\tau_a(t)$ is the proper time elapsed along the worldline of particle $a$. The functions $z_a(t)$ are refered to as the \textit{redshift} variables, \textit{e.g.} \cite{De.08,Sa.al.08,Bl.al.10,Bl.al2.10,BiDa.14,Bl.al.14,Bl.al2.14,Sh.al.14,BiDa.15,Jo.al.15}. The proof of Eq.~\eqref{z_a} involves three main steps: (i) the construction of a Fokker Lagrangian \cite{Fo.29} for the binary system by eliminating the gravitational field degrees of freedom in the total (matter plus field) Lagrangian, (ii) a perturbative redefinition of the particle's coordinate positions to remove all accelerations and higher-order time derivatives from this Fokker Lagrangian \cite{DaSc.91}, (iii) a Legendre transform to obtain an ordinary Hamiltonian; see Sec.~III of Ref.~\cite{Bl.al.13} for more details. These steps are analogous to those followed to compute, in the context of the post-Newtonian aproximation, an ordinary (Fokker-type) Hamiltonian for two point masses.\footnote{After imposing the usual ADM-TT gauge conditions, the field variables $h_{ij}^\text{TT}$ and $\dot{h}_{ij}^\text{TT}$ are eliminated from a Routh functional $R\bigl(\bo{x}_a,\bo{p}_a,h_{ij}^\text{TT},\dot{h}_{ij}^\text{TT}\bigr)$ constructed from the total (matter plus field) ADM Hamiltonian, yielding a (matter only) higher-order Hamiltonian $\widetilde{H}(\bo{x}_a,\bo{p}_a,\dot{\bo{x}}_a,\dot{\bo{p}}_a,\cdots)$ that can be reduced to an ordinary Hamiltonian $H(\bo{x}'_a,\bo{p}'_a)$ through a canonical transformation; see \textit{e.g.} Refs.~\cite{JaSc.98,Da.al.00,JaSc.12}.}

If the changes $\delta r(t)$, $\delta p_r(t)$ and $\delta L$ map two neighboring solutions of the binary dynamics, then Hamilton's equations \eqref{EOM} must be satisfied, and we obtain
\beq\label{dM}
	\delta M = - \dot{p}_r \, \delta r + \dot{r} \, \delta p_r + \dot{\varphi} \, \delta L + \sum_a z_a \, \delta m_a \, ,
\eeq
where we used Eq.~\eqref{z_a} and the fact that, ``on shell,'' the Hamiltonian is numerically equal to the ADM mass $M$. Since $\dot{r}(t)$, $\dot{\varphi}(t)$, $\dot{p}_r(t)$, $z_a(t)$, $\delta r(t)$ and $\delta p_r(t)$ are not constant, the relation \eqref{dM} is not particularly useful.

However, we may get rid of these orbital variations by taking the \textit{average} of Eq.\,\eqref{dM} over one radial period of the motion. Introducing the notation $\av{f} \equiv \frac{1}{P} \int_0^P f(t) \, \ud t$ for the time average of any function $f$ (recall that $P$ is the coordinate time period of the radial motion), and using the fact that $M$, $L$ and $m_a$ are constants of the motion, we get
\beq\label{<dM>}
	\delta M = \av{\dot{\varphi}} \, \delta L + \av{\dot{r} \, \delta p_r - \dot{p}_r \, \delta r} + \sum_a \av{z_a} \, \delta m_a \, .
\eeq
This variational relationship is clearly reminiscent of the first law of mechanics established in Paper I for circular motion [see Eq.~(1.1) there], with the constant circular-orbit frequency and the constant redshifts replaced by their orbit-averaged counterparts $\av{\dot{\varphi}}$ and $\av{z_a}$. These averaged quantities carry clear physical interpretations. Indeed,
\begin{subequations}
	\begin{align}
		\av{\dot{\varphi}} &= \frac{1}{P} \int_0^P \! \dot{\varphi}(t) \, \ud t = \frac{1}{P} \int_0^\Phi \! \ud \varphi = \frac{\Phi}{P} \, , \\
		\av{z_a} &= \frac{1}{P} \int_0^P \! \dot{\tau}_a(t) \, \ud t = \frac{1}{P} \int_0^{T_a} \! \ud \tau_a = \frac{T_a}{P} \, , \label{<zA>}
	\end{align}
\end{subequations}
where $\Phi \equiv 2\pi + \Delta \Phi$ is the accumulated azimuthal angle (or orbital phase) per radial period, with $\Delta \Phi$ the relativistic periastron advance, and $T_a$ is the proper time period of the radial motion of particle $a$. Notice that $P$ also coincides with the proper time period of the radial motion for a distant inertial observer, where the geometry is essentially flat.

The radial contribution in Eq.~\eqref{<dM>}, which vanishes for circular motion, can also be given a simple physical interpretation. Integrating by parts and using the periodicity of $p_r(t) \, \delta r(t)$, with period $P$, we obtain
\beq\label{radial}
	\av{\dot{r} \, \delta p_r - \dot{p}_r \, \delta r} = \av{\dot{r} \, \delta p_r + p_r \, \delta \dot{r}} = \av{\delta(\dot{r} \, p_r)} = n \, \delta R \, ,
\eeq
where $n \equiv 2\pi / P$ is the radial frequency and $R \equiv \frac{1}{2\pi} \oint p_r \, \ud r = \frac{2}{2\pi} \int_{r_\text{min}}^{r_\text{max}} p_r \, \ud r$ the radial action integral. (The last step in Eq.~\eqref{radial} made use of the change of variable $t \to r$ in the integral.) Finally, introducing the notation $\omega \equiv \av{\dot{\varphi}}$ for the average azimuthal frequency, we obtain the variational relationship
\beq\label{first_law}
	\delta M = \omega \, \delta L + n \, \delta R + \sum_a \av{z_a} \, \delta m_a \, .
\eeq
This is our first law of mechanics for binary systems of nonspinning compact objects moving along nearby eccentric orbits. Equation \eqref{first_law} generalizes to generic (bound) orbits the first law of Paper I, previously established for circular motion. In order to emphasize the structure (time period or angular period) $\times$ (variation of the ``conjugate variable''), Eq.~\eqref{first_law} may also be written as
\beq
	P \, \delta M = \Phi \, \delta L + 2 \pi \, \delta R + \sum_a T_a \, \delta m_a \, .
\eeq
The occurrence of the action integrals $L = \frac{1}{2\pi} \oint p_\varphi \ud \varphi$ and $R = \frac{1}{2\pi} \oint p_r \ud r$ suggests an alternative proof of the first law of binary mechanics, based on the use of action-angle variables.

\subsection{First law from action-angle variables}\label{subsec:q_J}

In the previous section, we studied a 4-dimensional dynamical system with a Hamiltonian $H(r,\varphi,p_r,p_\varphi;m_a)$ that depends on two external parameters $m_a$. In this section, we shall now consider an extended, 8-dimensional phase space with 4 generalized coordinates $(r,\varphi,\tau_1,\tau_2)$ and 4 conjugate momenta $(p_r,p_\varphi,p_{\tau_1},p_{\tau_2})$. In particular, the proper times $\tau_a$ elapsed along the worldlines of the particles now play the role of two additional (non-compact) coordinates. Therefore, in addition to Eqs.~\eqref{EOM} we have the extra canonical equations
\beq\label{tau}
	\dot{\tau}_a = \frac{\partial H}{\partial p_{\tau_a}} \, , \quad \dot{p}_{\tau_a} = - \frac{\partial H}{\partial \tau_a} \, .
\eeq
Since the Hamiltonian does not depend explicitly on the proper times $\tau_a$, the momenta $p_{\tau_a}$ must be constants of the motion. Comparing \eqref{tau} with \eqref{z_a}, it is clear that $p_{\tau_a} = m_a$.

Now, this 8-dimensional dynamical system possesses 4 first integrals $P_\alpha \equiv (M,L,m_1,m_2)$ that are independent (for non-degenerate orbits) and in involution, i.e., which have vanishing Poisson brackets.\footnote{This is obvious from the triviality of the first integrals $P_\alpha = (H,p_\varphi,p_{\tau_1},p_{\tau_2})$.} Therefore, this dynamical system is \textit{completely integrable} \cite{Arn}. For bound orbits, the motion in phase space is bounded in the $r$ and $\varphi$ directions; however it is not bounded in the $\tau_a$ directions, such that the Liouville-Arnol'd theorem does not guarantee the existence of action-angle type variables \cite{Arn}. Nevertheless, a generalization of this theorem by Fiorani \textit{et al.} \cite{Fi.al.03} ensures the existence of \textit{generalized} action-angle variables $(q_\alpha,J_\alpha)$ for completely integrable dynamical systems with non-compact level sets; see Secs. II A and B of Ref.~\cite{HiFl.08} for a nice summary, given in the language of symplectic geometry.

Thus, for our dynamical system we can define the generalized action variables \cite{Fi.al.03,HiFl.08}
\begin{subequations}\label{J_alpha}
	\begin{align}
		J_r &\equiv \frac{1}{2\pi} \oint p_r \, \ud r = R \, , \label{J_r} \\
		J_\varphi &\equiv \frac{1}{2\pi} \oint p_\varphi \, \ud \varphi = L \, , \\
		J_{\tau_a} &\equiv \frac{1}{2\pi} \int_0^{2\pi} \! p_{\tau_a} \, \ud \tau_a = m_a \, .
	\end{align}
\end{subequations}
Since the Hamiltonian $H(r,p_r,L,m_a) = M$ does not depend on the coordinates $\varphi$, $\tau_1$ and $\tau_2$, the radial momentum $p_r$ can be expressed solely as a function of $r$ and the first integrals $P_\alpha$. Hence the action variables $J_\alpha = (J_r,J_\varphi,J_{\tau_1},J_{\tau_2})$ are functions of the first integrals only. The generalization of the Liouville-Arnol'd theorem for non-compact level sets \cite{Fi.al.03} guarantees that these relationships can be inverted, yielding the expressions $P_\alpha(J_\beta)$ of the first integrals in terms of the action variables.

Then, the complete solution of the Hamilton-Jacobi equation for the action $S(t,s_\alpha,P_\alpha) = - M \, t + W(s_\alpha;P_\alpha)$ of the two-body system can easily be found by separation of the variables $s_\alpha \equiv (r,\varphi,\tau_1,\tau_2)$, yielding the following expression for Hamilton's characteristic function:
\beq\label{S_0}
	W(s_\alpha;P_\alpha) = \int^r \! p_r(r';P_\alpha) \, \ud r' + L \, \varphi + \sum_a m_a \tau_a \, .
\eeq
From the characteristic function $W$ we define a generating function $G(s_\alpha,J_\alpha) \equiv W(s_\alpha;P_\alpha(J_\beta))$ that generates a Type II canonical transformation, yielding a new Hamiltonian $H'$ expressed in generalized action-angle variables $(q_\alpha,J_\alpha)$. Since $G$ does not depend explicitly on time $t$, we have $H'(q_\alpha,J_\alpha) = H(s_\alpha,p_\alpha)$. The generalized angle variables can be computed from the relation $q_\alpha = \partial G / \partial J_\alpha$. On the other hand, $p_\alpha = \partial G / \partial s_\alpha$ is already satisfied from Eq.~\eqref{S_0}. In generalized action-angle variables, Hamilton's equations take on the simple form
\beq\label{Jdot}
	\dot{q}_\alpha = \frac{\partial H'}{\partial J_\alpha} \equiv \omega_\alpha \, , \quad \dot{J}_\alpha = - \frac{\partial H'}{\partial q_\alpha} = 0 \, .
\eeq
Because the Hamiltonian $H'(J_\alpha)$ does not depend on the generalized angles $q_\alpha(t)$, the actions $J_\alpha$ are constants of the motion, as expected from the fact that they are functions of the first integrals $P_\alpha$ only. Thus the \textit{fundamental frequencies} $\omega_\alpha$ are also constants of the motion, such that $q_\alpha(t) = q_{\alpha,0} + \omega_\alpha t$, where $q_{\alpha,0}$ are four constants that reflect a choice of ``initial phase.'' The wide class of coordinate transformations that leave the $\omega_\alpha$ unchanged is discussed, \textit{e.g.}, in Ref.~\cite{HiFl.08} (see also \cite{Sc.02,BaSa.11}). While $\omega_r$ and $\omega_\varphi$ are true angular frequencies, the quantities $\omega_{\tau_1}$ and $\omega_{\tau_2}$ are dimensionless.

Now, by varying the Hamiltonian $H'$ and using the equations of motion \eqref{Jdot}, we immediately get $\delta H' = \sum_\alpha \omega_\alpha \, \delta J_\alpha$. Then, using Eqs.~\eqref{J_alpha} and the fact that $H' = M$ ``on shell,'' we obtain the variational relation
\beq
	\delta M = \omega_r \, \delta R + \omega_\varphi \, \delta L + \sum_a \omega_{\tau_a} \, \delta m_a \, .
\eeq
Moreover, since the fundamental frequencies associated with the (generalized) angle variables $q_r$, $q_\varphi$ and $q_{\tau_a}$ read $\omega_r = 2\pi / P = n$, $\omega_\varphi = \Phi / P = \omega$ and $\omega_{\tau_a} = T_a / P = \av{z_a}$, we recover the first law \eqref{first_law}. Notice also that the averaging over one radial period is already taken care of when using action-angle type variables.

Finally, we discuss the special case of circular orbits, which are characterize by a vanishing radial action integral: $R = 0$. For such \textit{degenerate} orbits, the constants of the motion $M$ and $L$ are no longer independent, and the motion becomes simply periodic with constant angular frequency $\omega$. In that limit, an interesting coordinate-invariant relationship is given by the reduced periastron advance $K \equiv \omega / n = \Phi / (2\pi)$ as a function of the circular-orbit frequency $\omega$ (see, \textit{e.g.}, Refs.~\cite{Da.al.00,Ba.al.10,Mr.al.10,BaSa.11,Le.al.11,Fo.al.13,Le.al.13,Hi.al.13}), computed from the zero-eccentricity limit of $K = - \partial R / \partial L$, evaluated at fixed $M$, $m_1$ and $m_2$.

\section{Consequences of the first law}\label{sec:consequences}

We now explore some straighforward consequences of the first law for eccentric-orbit binaries. In particular, in section \ref{subsec:PDE} we derive various partial differential equations relating the coordinate-invariant relationships $M(\omega,n,m_a)$, $L(\omega,n,m_a)$, $R(\omega,n,m_a)$ and $\av{z_a}(\omega,n,m_a)$, and we establish a ``first integral'' associated with the variational law \eqref{first_law} in section \ref{subsec:1st_int}.

\subsection{Partial differential equations}\label{subsec:PDE}

For given particles' masses $m_a$, the action integrals $L$ and $R$ uniquely specify an orbit up to initial conditions. Using $(L,R,m_1,m_2)$ as independent variables, the variational first law \eqref{first_law} thus yields the following set of partial differential equations:
\beq\label{blob}
	\frac{\partial M}{\partial L} = \omega \, , \quad \frac{\partial M}{\partial R} = n \, , \quad \frac{\partial M}{\partial m_a} = \av{z_a} \, .
\eeq
In words, the partial derivatives of the ADM mass $M$ with respect to the generalized action variables $L$, $R$, $m_1$ and $m_2$ are simply given by the fundamental frequencies $\omega$, $n$, $\av{z_1}$ and $\av{z_2}$. However, it is often more convenient to parameterize the motion using the two frequencies $\omega$ and $n$ instead of the action integrals $L$ and $R$. Using $(\omega,n,m_1,m_2)$ as independent variables, Eq.~\eqref{first_law} is equivalent to the relationships
\begin{subequations}\label{PDEs}
	\begin{align}
		\frac{\partial M}{\partial \omega} &= \omega \frac{\partial L}{\partial \omega} + n \frac{\partial R}{\partial \omega} \, , \label{PDEOmega} \\
		\frac{\partial M}{\partial n} &= \omega \frac{\partial L}{\partial n} + n \frac{\partial R}{\partial n} \, , \label{PDEn} \\
		\frac{\partial M}{\partial m_a} &= \omega \frac{\partial L}{\partial m_a} + n \frac{\partial R}{\partial m_a} + \av{z_a} \, . \label{PDEm}
	\end{align}
\end{subequations}
By taking partial derivatives of \eqref{PDEm} with respect to $m_1$ and $m_2$, and using the commutation of partial derivatives, we obtain the simple result
\beq
	\frac{\partial \av{z_1}}{\partial m_2} = \frac{\partial \av{z_2}}{\partial m_1} \, ,
\eeq
which can be seen as reflecting some \textit{average equilibrium} state of the binary under the mutual gravitational attraction of its components. This generalizes a similar result established for circular motion in Paper I [see Eq.~(2.42) there]. 

Next, by taking partial derivatives of Eqs.~\eqref{PDEs}, we shall derive relationships between the functions $M(\omega,n,m_a)$, $L(\omega,n,m_a)$ and $\av{z_a}(\omega,n,m_a)$, while eliminating $R(\omega,n,m_a)$. First, we can express $\partial^2 R / \partial \omega \partial m_a$ in two independent ways by computing the derivative of \eqref{PDEOmega} with respect to $m_a$ and the derivative of \eqref{PDEm} with respect to $\omega$, yielding
\beq\label{dLdm}
	\frac{\partial L}{\partial m_a} = - \frac{\partial \av{z_a}}{\partial \omega} \, .
\eeq
Similarly, we can express $\partial^2 R / \partial n \partial m_a$ in two ways by computing the derivative of Eq.~\eqref{PDEn} with respect to $m_a$ and the derivative of Eq.~\eqref{PDEm} with respect to $n$. Combined with \eqref{dLdm}, the resulting expression yields
\beq\label{dMdm}
	\frac{\partial M}{\partial m_a} = \av{z_a} - \omega \frac{\partial \av{z_a}}{\partial \omega} - n \frac{\partial \av{z_a}}{\partial n} \, .
\eeq
The relationships \eqref{dLdm} and \eqref{dMdm} generalize to eccentric orbits Eqs.~(4.7) and (4.9) of Paper I. These results should prove useful when applied in the context of black hole perturbation theory. Indeed, existing gravitational self-force computations already provide numerical data for the invariant function $\av{z_1}(\omega,n)$ for a particle of mass $m_1$ on a generic bound (eccentric) orbit around a Schwarzschild black hole of mass $m_2 \gg m_1$ \cite{BaSa.11,Ak.al.15,vMeSh.15}. Combining such results with Eqs.~\eqref{dLdm} and \eqref{dMdm}, crucial information regarding the total energy $M(\omega,n)$ and orbital angular momentum $L(\omega,n)$ of the \textit{binary} system could be inferred, similarly to what has been done in Refs.~\cite{Le.al.12,Le.al2.12,Ba.al.12} for circular motion; see Secs.~\ref{subsec:GSF} and \ref{subsec:E-L-R} below. 

Finally, we can express $\partial^2 R / \partial \omega \partial n$ in two ways by computing the derivative of Eq.~\eqref{PDEOmega} with respect to $n$ and the derivative of Eq.~\eqref{PDEn} with respect to $\omega$, yielding
\beq\label{dMdOmega}
	\frac{\partial M}{\partial \omega} = \omega \frac{\partial L}{\partial \omega} + n \frac{\partial L}{\partial n} \, .
\eeq
This expression extends to eccentric orbits the so-called thermodynamic relation commonly used in PN theory for quasi-circular orbits (see, \textit{e.g.}, Refs.~\cite{Da.al.00,Bl.02}), or in the construction of sequences of quasi-equilibrium initial data for binary black holes and neutrons stars \cite{Go.al.02,Gr.al.02,Sh.al.04,Ca.al2.06,Ts.al.15}.

Finally, by combining Eqs.~\eqref{PDEm}, \eqref{dLdm} and \eqref{dMdm}, we can relate the rates of change of the radial action with respect to the particles' masses to the rates of change of the averaged redshifts with respect to the radial frequency via
\beq\label{dRdm}
	\frac{\partial R}{\partial m_a} = - \frac{\partial \av{z_a}}{\partial n} \, .
\eeq

\subsection{First integral relation}\label{subsec:1st_int}

Since Einstein's equation does not involve any privileged mass scale, the ADM mass $M$ must be a homogeneous function of degree 1 in the four variables $L^{1/2}$, $R^{1/2}$, $m_1$ and $m_2$, i.e., $M(\lambda L^{1/2}, \lambda R^{1/2}, \lambda m_1, \lambda m_2) = \lambda \, M(L^{1/2}, R^{1/2}, m_1, m_2)$ for any $\lambda \neq 0$. Therefore, applying Euler's theorem together with Eqs.~\eqref{blob} immediately yields the first integral relation
\beq\label{first_integral}
	M = 2 (\omega L + n R) + \sum_a m_a \av{z_a} \, .
\eeq
This generalizes a similar result established for circular motion in Paper I. Loosely speaking, the total mass-energy $M$ is given by the sum of a ``azimuthal energy'' $2\omega L$, a ``radial energy'' $2 n R$, and the redshifted masses $m_a \av{z_a}$. Since Einstein's equation is nonlinear, it is remarkable that the total mass-energy of the binary system can be written in such a simple way in terms of the individual masses of the particles and other invariant quantities characterizing the orbit.

Alternatively, using the time average introduced in Sec.~\ref{subsec:average} above, the first integral \eqref{first_integral} can be recast in the form
\beq
	M = \Big\langle 2 \, (\dot{r} \, p_r + \dot{\varphi} \, p_\varphi) + \sum_a m_a z_a \Big\rangle \, .
\eeq
However, the different terms in the right-hand side do not possess simple physical interpretations in terms of kinetic energy $E_\text{k}$ and gravitational potential energy $U$. Indeed, in the Newtonian limit $c^{-1} \to 0$, the first term gives $2 \av{\dot{r} \, p_r + \dot{\varphi} \, p_\varphi} = 4 \av{E_\text{k}}$ while the second term yields $\sum_a m_a \av{z_a} = m - \av{E_\text{k}} + 2\av{U}$. The Newtonian virial theorem implies $\av{U} = - 2 \av{E_\text{k}}$, and we recover the total energy $M = m - \av{E_\text{k}} = m + \av{U}/2$ of a Keplerian elliptic orbit.

\section{Verification of the first law}\label{sec:verification}

By making use of available results for the conservative PN dynamics of two point masses, we shall now check that the first law of binary mechanics \eqref{first_law} and the first integral relation \eqref{first_integral} are indeed satisfied up to 3PN order.

We use the result of Ref.~\cite{Da.al.00} for the 3PN-accurate expression of the radial action $R(\hat{E},\hat{L})$ as a function of the reduced binding energy $\hat{E} \equiv (M - m) / \mu$ and the dimensionless angular momentum $\hat{L} \equiv L / (m \mu)$, where $m = m_1 + m_2$ is the total mass and $\mu = m_1 m_2 / m$ is the reduced mass. Following Ref.~\cite{Ar.al.08}, we choose instead to parameterize the orbit in terms of the dimensionless variables
\beq\label{epsilon_j}
	\varepsilon \equiv - 2 \hat{E} \, , \quad j \equiv - 2 \hat{E} \hat{L}^2 \, ,
\eeq
such that in the Newtonian limit $\varepsilon \sim m / a$ and $j \sim 1 - e^2$, where $a$ and $e$ are the semi-major axis and eccentricity of a Keplerian orbit. In terms of the variables $\varepsilon$ and $j$, the ADM mass, orbital angular momentum and radial action are given by \cite{Da.al.00,Da.al2.00,Da.al.01}
\begin{subequations}\label{M_L_R}
	\begin{align}
		M &= m - \frac{m \nu}{2} \, \varepsilon \, , \\
		L &= m^2 \nu \, \sqrt{\frac{j}{\varepsilon}} \, , \\
		R &= \frac{m^2 \nu}{\sqrt{\varepsilon}} \, \biggl\{ 1 - \sqrt{j} + \biggl( - \frac{15}{8} + \frac{\nu}{8} + \frac{3}{\sqrt{j}} \biggr) \, \varepsilon \nonumber \\ &+ \biggl( \frac{35}{128} + \frac{15}{64} \nu + \frac{3}{128} \nu^2 - \biggl[ \frac{15}{4} - \frac{3}{2} \nu \biggr] \frac{1}{\sqrt{j}} + \biggl[ \frac{35}{4} - \frac{5}{2} \nu \biggr] \frac{1}{j^{3/2}} \biggr) \, \varepsilon^2 \label{R} \\ &+ \biggl( \frac{21}{1024} - \frac{105}{1024} \nu + \frac{15}{1024} \nu^2 + \frac{5}{1024} \nu^4 - \biggl[ \frac{105}{4} - \frac{109}{3} \nu + \frac{41}{128} \pi^2 \nu + \frac{15}{4} \nu^2 \biggr] \frac{1}{j^{3/2}} \nonumber \\ &\qquad + \biggl[ \frac{15}{16} - \frac{15}{16} \nu + \frac{3}{4} \nu^2 \biggr] \frac{1}{\sqrt{j}} + \biggl[ \frac{231}{4} - \frac{125}{2} \nu + \frac{123}{128} \pi^2 \nu + \frac{21}{8} \nu^2 \biggr] \frac{1}{j^{5/2}} \biggr) \, \varepsilon^3 + o(\varepsilon^3) \biggr\} \, , \nonumber
	\end{align}
\end{subequations}
where $\nu \equiv \mu / m = m_1 m_2 / m^2$ is the symmetric mass ratio, such that $\nu = 1/4$ for equal-mass binaries and $\nu \to 0$ in the extreme mass-ratio limit.

The observables of the orbital motion, namely the periastron-to-periastron period $P$ and the angular advance per radial period $\Phi$ then follow from the first two relations in Eqs. \eqref{blob}, which can be combined to give $P/2\pi = \partial R / \partial M \vert_L$ and $\Phi/2\pi = - \partial R / \partial L \vert_M$. By computing the ratio of these expressions we recover $\omega  = \Phi/P = \partial M / \partial L \vert_R$. Performing the change of variables $(M,L) \to (\varepsilon,j)$ and using the chain rule while computing the partial derivatives of \eqref{R}, we obtain the following 3PN-accurate expansions for the frequencies of the motion:
\begin{subequations}\label{n_omega}
	\begin{align}
		n = \frac{\varepsilon^{3/2}}{m} \, \bigg\{ 1 &+ \biggl( - \frac{15}{8} + \frac{\nu}{8} \biggr) \, \varepsilon \nonumber \\ &+ \biggl( \frac{555}{128} + \frac{15}{64} \nu + \frac{11}{128} \nu^2 - \biggl[ \frac{15}{2} - 3 \nu \biggr] \frac{1}{\sqrt{j}} \biggr) \, \varepsilon^2 \nonumber \\ &+ \biggl( - \frac{9795}{1024} - \frac{1665}{1024} \nu - \frac{105}{1024} \nu^2 + \frac{45}{1024} \nu^3 + \biggl[ \frac{255}{8} - \frac{135}{8} \nu + \frac{15}{4} \nu^2 \biggr] \frac{1}{\sqrt{j}} \nonumber \\ &\qquad - \biggl[ \frac{105}{2} - \frac{218}{3} \nu + \frac{41}{64} \pi^2 \nu + \frac{15}{2} \nu^2 \biggr] \frac{1}{j^{3/2}} \biggr) \, \varepsilon^3 + o(\varepsilon^3) \bigg\} \, , \label{n} \\
		\omega = \frac{\varepsilon^{3/2}}{m} \, \bigg\{ 1 &+ \biggl( - \frac{15}{8} + \frac{\nu}{8} + \frac{3}{j} \biggr) \, \varepsilon \nonumber \\ &+ \biggl( \frac{555}{128} + \frac{15}{64} \nu + \frac{11}{128} \nu^2 - \biggl[ \frac{15}{2} - 3 \nu \biggr] \frac{1}{\sqrt{j}} - \biggl[ \frac{75}{8} - \frac{15}{8} \nu \biggr] \frac{1}{j} + \biggl[ \frac{105}{4} - \frac{15}{2} \nu \biggr] \frac{1}{j^2} \biggr) \, \varepsilon^2 \nonumber \\ &+ \biggl( - \frac{9795}{1024} - \frac{1665}{1024} \nu - \frac{105}{1024} \nu^2 + \frac{45}{1024} \nu^3 + \biggl[ \frac{255}{8} - \frac{135}{8} \nu + \frac{15}{4} \nu^2 \biggr] \frac{1}{\sqrt{j}} \nonumber \\ &\qquad + \biggl[ \frac{2685}{128} - \frac{225}{64} \nu + \frac{153}{128} \nu^2 \biggr] \frac{1}{j} - \biggl[ 75 - \frac{245}{3} \nu + \frac{41}{64} \pi^2 \nu + \frac{15}{2} \nu^2 \biggr] \frac{1}{j^{3/2}} \nonumber \\ &\qquad - \biggl[ \frac{4095}{32} - \frac{4043}{32} \nu + \frac{123}{128} \pi^2 \nu + \frac{195}{16} \nu^2 \biggr] \frac{1}{j^2} \nonumber \\ &\qquad + \biggl[ \frac{1155}{4} - \frac{625}{2} \nu + \frac{615}{128} \pi^2 \nu + \frac{105}{8} \nu^2 \biggr] \frac{1}{j^3} \biggr) \, \varepsilon^3 + o(\varepsilon^3) \bigg\} \, . \label{omega}
	\end{align}
\end{subequations}
Equation \eqref{n} agrees with Eq.~(7.7a) of Ref.~\cite{Ar.al.08}, while Eq.~\eqref{omega} can easily be recovered by combining their Eqs.~(7.7a) and (7.7b). At Newtonian order, we have $\omega = n$, or equivalently $\Delta \Phi = 0$, which is to say no periastron advance. As is well known, bound Keplerian orbits are closed ellipses.

The formulas \eqref{R} and \eqref{n_omega} were derived from the expression for the 3PN binary Hamiltonian, in the center-of-mass frame, in ADM-TT coordinates \cite{Da.al.00}. The coordinate-invariant relations $\av{z_a}(\varepsilon,j)$ were, on the other hand, recently computed, up to 3PN order, by following an entirely different route. Indeed, the near-zone metric $g_{\alpha\beta}(y_1) \equiv g_{\alpha\beta}(t,\bo{y}_1)$ evaluated at the coordinate location $\bo{y}_1(t)$ of the particle $1$ was computed up to 3PN order, in harmonic coordinates \cite{Bl.al.10}. Akcay \textit{et al.} \cite{Ak.al.15} then obtained the redshift $z_1 = (- g_{\alpha\beta}(y_1) \, v_1^\alpha v_1^\beta)^{1/2}$ of this particle by contracting the near-zone metric with the coordinate velocity $v_1^\alpha \equiv \ud y_1^\alpha / \ud t = (1,v_1^i)$. Specializing the resulting expression to the center-of-mass frame, using the 3PN generalized quasi-Keplerian representation of the motion \cite{Me.al2.04}, and performing an average over one radial period, they obtained the following 3PN-accurate expression for the average redshift:\footnote{Reference \cite{Ak.al.15} computed instead the average of $u_1^t = 1 / z_1$ with respect to \textit{proper} time (denoted $\av{U}$ there); we simply have $\av{z_1} = 1 / \av{U}$.}
\begin{align}\label{<z>}
	{\langle z_1 \rangle} = 1 &+ \biggl( - \frac{3}{4} - \frac{3}{4} \Delta + \frac{\nu}{2} \biggr) \, \varepsilon \nonumber \\ &+ \biggl( \frac{15}{8} + \frac{15}{8} \Delta - \frac{3}{16} \nu - \frac{3}{16} \Delta \, \nu + \frac{\nu^2}{4} - \frac{3 + 3 \Delta }{\sqrt{j}} \biggr) \, \varepsilon^2 \nonumber \\ &+ \biggl( - \frac{65}{16} - \frac{65}{16} \Delta - \frac{3}{32} \nu^2 - \frac{3}{32} \Delta \, \nu^2 + \frac{\nu^3}{8} \nonumber \\ &\qquad + \biggl[ \frac{105}{8} + \frac{105}{8} \Delta - \frac{33}{8} \nu - \frac{33}{8} \Delta \, \nu + 3 \nu^2 \biggr] \frac{1}{\sqrt{j}} \nonumber \\ &\qquad - \biggl[ \frac{35}{2} + \frac{35}{2} \Delta - \frac{25}{4} \nu - \frac{25}{4} \Delta \, \nu + 5 \nu^2 \biggr] \frac{1}{j^{3/2}} \biggr) \, \varepsilon^3 \nonumber \\ &+ \biggl( \frac{291}{32} + \frac{291}{32} \Delta + \frac{65}{64} \nu + \frac{65}{64} \Delta \, \nu + \frac{15}{128} \nu^2 + \frac{15}{128} \Delta \, \nu^2 - \frac{3}{64} \nu^3 - \frac{3}{64} \Delta \, \nu^3 + \frac{\nu^4}{16} \nonumber \\ &\qquad - \biggl[ \frac{5625}{128} + \frac{5625}{128} \Delta - \frac{1125}{64} \nu - \frac{1125}{64} \Delta \, \nu + \frac{1749}{128} \nu^2 + \frac{549}{128} \Delta \, \nu^2 - \frac{33}{8} \nu^3 \biggr] \frac{1}{\sqrt{j}} \nonumber \\ &\qquad + \biggl[ \frac{45}{2} + \frac{45}{2} \Delta - 9 \nu - 9 \Delta \, \nu \biggr] \frac{1}{j} + \biggl[ \frac{1785}{16} + \frac{1785}{16} \Delta + \left( - \frac{13543}{96} + \frac{287}{256} \pi^2 \right) \nu \nonumber \\ &\qquad\qquad + \left( - \frac{13543}{96} + \frac{287}{256} \pi^2 \right) \Delta \, \nu + \left( \frac{9391}{96} - \frac{41}{64} \pi^2 \right) \nu^2 + \frac{505}{32} \Delta \, \nu^2 - \frac{125}{8} \nu^3 \biggr] \frac{1}{j^{3/2}} \nonumber \\ &\qquad - \biggl[ \frac{693}{4} + \frac{693}{4} \Delta + \left( - \frac{875}{4} + \frac{861}{256} \pi^2 \right) \nu + \left( - \frac{875}{4} + \frac{861}{256} \pi^2 \right) \Delta \, \nu \nonumber \\ &\qquad\qquad + \left( \frac{271}{2} - \frac{123}{64} \pi^2 \right) \nu^2 + \frac{21}{2} \Delta \, \nu^2 - \frac{21}{2} \nu^3 \biggr] \frac{1}{j^{5/2}} \biggr) \, \varepsilon^4 + o(\varepsilon^4) \, .
\end{align}
Here, $\Delta \equiv (m_2 - m_1) / m = (1-4\nu)^{1/2}$ is the reduced mass difference. (We assume $m_1 \leq m_2$.) The expression for $\av{z_2}(\varepsilon,j)$ is easily found by setting $\Delta \longrightarrow - \Delta$ in Eq.~\eqref{<z>}.

Now it is straightforward to check that the 3PN results \eqref{M_L_R}--\eqref{<z>} do indeed obey the first integral relation \eqref{first_integral}, as well as the partial differential equations \eqref{PDEs} with the substitutions $(\partial / \partial \omega, \partial / \partial n) \longrightarrow (\partial / \partial \varepsilon, \partial / \partial j)$. This verification provides a powerful check of the intricate calculations that resulted in the expression for the 3PN Hamiltonian in ADM-TT coordinates \cite{Da.al.00}, of the 3PN near-zone metric in harmonic coordinates \cite{Bl.al.10}, and of the first law itselfy. Moreover, if the redshift $\av{z_1}$ were computed up to 4PN order, the verification of Eq.~\eqref{first_law} would provide a powerful test of the recently derived 4PN binary Hamiltonian \cite{JaSc.12,JaSc.13,Da.al.14}.

Finally, we discuss the reduction to the case of circular orbits; degenerate orbits for which the frequencies $\omega$ and $n$ are not independent. By definition, a circular orbit has a vanishing radial action integral. Setting $R = 0$ in Eq.~\eqref{R} yields a relationship between $\varepsilon$ and $j$ (or equivalently between $M$ and $L$) that reads, up to 3PN order \cite{Da.al.00,Ar.al.08},
\begin{align}\label{j}
	j = 1 &+ \left( \frac{9}{4} + \frac{\nu}{3} \right) \varepsilon + \left( \frac{81}{16} - 2 \nu + \frac{\nu^2}{16} \right) \varepsilon^2 \nonumber \\ &+ \left( \frac{2835}{192} - \left[ \frac{7699}{392} - \frac{41}{32} \pi^2 \right] \nu + \frac{\nu^2}{2} + \frac{\nu^3}{64} \right) \varepsilon^3 + o(\varepsilon^3) \, .
\end{align}
Substituting for this $j$ into \eqref{omega}, and inverting the resulting formula, we recover the well-known 3PN-acurate expression for the total mass-energy $M$ as a function of the circular-orbit frequency $\omega$. By inserting this expansion into Eqs.~\eqref{j} and \eqref{<z>}, we also recover the known 3PN results for $L(\omega)$ and $z_1(\omega)$ for circular orbits; see, \textit{e.g.}, Eqs.~(2.35)--(2.37) of Ref.~\cite{Le.al.12}.

\section{Applications of the first law}\label{sec:applications}

In this section we discuss several applications of the first law of binary mechanics, starting with the gravitational-wave driven adiabatic inspiral of compact binary systems (Sec.~\ref{subsec:adiabatic}). We then explain how this relation could be used, together with existing and/or forthcoming GSF calculations (Sec.~\ref{subsec:GSF}), to get strong-field information about the conservative dynamics of compact binaries, especially regarding the binding energy, angular momentum and radial action of such systems (Secs.~\ref{subsec:E-L-R} and \ref{subsec:Schw}). Finally, we illustrate how our results can be used to inform the conservative part of the EOB dynamics beyond circular motion (Sec.~\ref{subsec:EOB}).

\subsection{Gravitational-wave driven adiabatic evolution}\label{subsec:adiabatic}

The variational first law relates the total energy, orbital angular momentum, radial action integral, and averaged redshifts of \textit{two} physically distinct binary systems under small changes of the orbital frequencies $\omega,n$ and of the particles' masses $m_1,m_2$. In the problem of building template waveforms for inspiralling compact-object binaries, we wish to follow the evolution of a \textit{single} system as it gradually inspirals under the effect of gravitational radiation reaction. In this case, the quantities $M$, $L$, $R$ and $m_a$ are no longer constants of the motion.

Nevertheless, if the characteristic timescale $T_\text{r.r.}$ of gravitational radiation reaction is much larger than the typical orbital timescale $P$, then the time evolution of $M(t)$, $L(t)$, $R(t)$ and $m_a(t)$ is well approximated as an \textit{adiabatic} process. We may then identify the two physically distinct systems compared in the first law with two nearby states of a single binary system. This approximation is commonly adopted while computing sequences of quasi-equilibrium initial data for binary systems of black holes and neutrons stars \cite{Go.al.02,Gr.al.02,Sh.al.04,Ca.al2.06,Ts.al.15}. The variations $\delta M$ and $\delta L$ appearing in Eq.~\eqref{first_law} are then interpreted as the secular changes in the mechanical energy and angular momentum of the binary during an interval $P \lesssim \delta t \ll T_\text{r.r.}$, and similarly for $\delta R$ and $\delta m_a$. Dividing \eqref{first_law} by $\delta t$ and averaging over a radial period, the average rates of change of $M(t)$, $L(t)$, $R(t)$ and $m_a(t)$ must then obey
\beq\label{Mdot}
	\av{\dot{M}} = \omega \, \av{\dot{L}} + n \, \av{\dot{R}} + \sum_a \, \av{z_a} \, \av{\dot{m}_a} \, .
\eeq
Notice that the frequencies $\omega$, $n$ and $\av{z_a}$ were factored out from the averaging because they are already averaged over one radial period and they do not vary over a dynamical timescale.

We may then appeal to the usual argument of balance of energy and angular momentum, $\av{\dot{M}} = - \mathcal{F}$ and $\av{\dot{L}} = - \mathcal{G}$, where the gravitational-wave fluxes of energy $\mathcal{F}$ and angular momentum $\mathcal{G}$ can be computed from the far-zone gravitational field \cite{Bl.96,Ar.al2.09}. These (heuristic) balance equations are motivated by the (exact) Bondi-Sachs mass-loss formula \cite{Bo.al.62,Sa.62}; see \textit{e.g.} Sec. II E of Paper I for a discussion. Moreover for a neutron star the conservation of the baryonic mass implies $\av{\dot{m}_a} = 0$, while for a black hole tidal heating \cite{TaPo.08,PoVl.10} is responsible for an increase in the irreducible mass, the average rate of which is given by the flux of energy $\mathcal{H}_a$ through the horizon: $\av{\dot{m}_a} = \mathcal{H}_a$ \cite{Al.01,Po2.04}. Therefore, Eq.~\eqref{Mdot} becomes
\beq\label{adiabatic}
	n \, \av{\dot{R}} = - \mathcal{F} + \omega \, \mathcal{G} - \sum_a \varepsilon_a \, \av{z_a} \, \mathcal{H}_a \, ,
\eeq
where $\varepsilon_a = 0$ for a neutron star and $\varepsilon_a = 1$ for a black hole.

For a nonspinning black hole, the tidally induced flux of energy $\mathcal{H}_a$ through the horizon is a small, 4PN effect relative to the leading-order (Newtonian) fluxes $\mathcal{F}$ and $\mathcal{G}$ \cite{PoSa.95}, and will thus be neglected here. Using the known expressions for the leading-order fluxes of energy and angular momentum \cite{PeMa.63,Pe.64}, we get for \textit{any} binary system of compact objects
\beq\label{dRdt}
	\av{\dot{R}} = - \frac{32}{5} \, m \nu^2 \, x^{7/2} \left\{ \frac{1}{{(1-e^2)}^{7/2}} \left( 1 + \frac{73}{24} e^2 + \frac{37}{96} e^4 \right) - \frac{1 + \frac{7}{8} e^2}{{(1 - e^2)}^2} \right\} + \calO(x^{9/2}) \, ,
\eeq
where $x \equiv (m \omega)^{2/3}$ is the usual, dimensionless, frequency-related PN parameter, and $e$ is the eccentricity parametrizing the Keplerian orbit. At this order of approximation, $j = 1 - e^2$ and $n = \omega$. Now let $f(e)$ denote the factor in curly brackets in Eq.~\eqref{dRdt}. It is positive for all $0 \leq e < 1$ and has the asymptotic behaviors $f(e) \sim \frac{11}{3} \, e^2$ and $f(e) \sim \frac{425}{768\sqrt{2}} \, (1-e)^{-7/2}$ in the limits where $e \to 0$ and $e \to 1$, respectively. Hence $\av{\dot{R}} \leq 0$ and we recover the known result that, in the weak-field regime, gravitational radiation reaction decreases the noncircularity of the orbit, all the more so if the eccentricity is large \cite{Pe.64}. Of course $\!\av{\dot{R}}$ vanishes in the limit $e \to 0$. For circular orbits, the ``adiabatic first law'' \eqref{adiabatic} shows that the fluxes of energy and angular momentum must be proportional: $\mathcal{F} = \omega \, \mathcal{G}$. By using available PN expressions for these fluxes, this relationship can be checked up to 3PN order \cite{Ar.al.08,Ar.al2.09}.

For completeness, we note that the average rates of change $\av{\dot{n}}$ and $\av{\dot{\omega}}$ of the frequencies entering Eq.~\eqref{Mdot} are given by
\begin{subequations}
	\begin{align}
		\av{\dot{n}} &= - \frac{\partial n}{\partial M} \, \mathcal{F} - \frac{\partial n}{\partial L} \, \mathcal{G} + \sum_a \frac{\partial n}{\partial m_a} \, \mathcal{H}_a \, , \\
		\av{\dot{\omega}} &= - \frac{\partial \omega}{\partial M} \, \mathcal{F} - \frac{\partial \omega}{\partial L} \, \mathcal{G} + \sum_a \frac{\partial \omega}{\partial m_a} \, \mathcal{H}_a \, ,
	\end{align}
\end{subequations}
where the partial derivatives with respect to $M$, $L$ and $m_a$ can easily be computed by means of the chain rule, together with Eqs.~\eqref{epsilon_j} and \eqref{n_omega}.

\subsection{Conservative dynamics beyond the geodesic approximation}\label{subsec:GSF}

The derivations of the first law of binary mechanics given in section \ref{sec:derivation} did not rely on any small-velocity or weak-field expansion. However, they both relied on several key assumptions that are typically met by PN spacetimes, such as (i) a well-defined conservative/dissipative split of the binary dynamics, (ii) the existence of an autonomous Hamiltonian describing the conservative dynamics, (iii) a point-particle description of the compact objects, and (iv) the implicit use of a regularization scheme to subtract off the divergent self-fields of the particles. Therefore the applicability of the formula \eqref{first_law} to modelling compact-object binaries in the strong-field regime, for which the PN approximation breaks down, is not guaranteed.

However, several results suggest that Eq.~\eqref{first_law} does indeed hold in this context. First, by making use of the circular-orbit first law of Paper I, the authors of Refs.~\cite{Le.al2.12,Ak.al.12} could recover the \textit{exact} (numerical) value of the shift of the Schwarzschild ISCO frequency induced by the conservative part of the GSF \cite{BaSa.09,BaSa.10}. Moreover, Eq.~\eqref{first_law} is \textit{formally identical} to the (nonspinning limit of the) first law-type relations of Refs.~\cite{Le.14,Is.al.15}, which were derived in the context of black hole perturbation theory. In particular, Isoyama \textit{et al.} \cite{Is.al.15} have devised a Hamiltonian formulation of the dynamics of a self-gravitating particle subject to the conservative GSF, for bound orbits in a Kerr background; the restriction of their first law relationship to circular equatorial orbits was used to compute the GSF-induced frequency shift of the Kerr ISCO \cite{Is.al.14}. Independently, Vines and Flanagan \cite{ViFl.15} have recently proved that, for generic stable bound orbits in a Schwarzschild background, the dynamics of a pointlike object subject to the conservative piece of the osculating-geodesic-sourced GSF \cite{PoPo.08} is Hamiltonian and integrable.

Thereafter, we shall thus assume that the first law \eqref{first_law} (with $\delta m_2 \to 0$) can be applied, together with perturbative GSF calculations, to obtain new information about the conservative dynamics of nonspinning compact binaries, even in a strong-field regime. Alternatively, one might adopt the viewpoint that the formula \eqref{first_law} \textit{defines}, in the perturbative context, some physically motivated notions of binding energy, angular momentum and radial action.

\subsection{Binding energy, angular momentum and radial action}\label{subsec:E-L-R}

Before we proceed to discuss these applications, let us first establish a few useful formulas. Introducing the variable $\calM \equiv M - \omega L - n R$, the relationships \eqref{dLdm}--\eqref{dRdm} can be combined to express $M$, $L$, $R$ and $\av{z_a}$ solely in terms of $\calM$ and its partial derivatives with respect to $\omega$, $n$ and $m_a$ as
\begin{subequations}
	\begin{align}
		M &= \calM - \omega \, \frac{\partial \calM}{\partial \omega} - n \, \frac{\partial \calM}{\partial n} \, , \label{M_calM} \\
		L &= - \frac{\partial \calM}{\partial \omega} \, , \\
		R &= - \frac{\partial \calM}{\partial n} \, , \label{R_calM} \\
		\av{z_a} &= \frac{\partial \calM}{\partial m_a} \, . \label{za_dcalMdma}
	\end{align}
\end{subequations}
Thereafter it will prove convenient to introduce the dimensionless quantities $\hat{E} \equiv (M - m) / \mu$, $\hat{L} \equiv L / (m \mu)$, $\hat{R} \equiv R / (m \mu)$ and $\hat{\calM} \equiv (\calM - m) / \mu$, and to perform the change of variables $(m_1,m_2,n,\omega) \longrightarrow (m,\nu,\hat{n},\hat{\omega})$, where $\hat{n} \equiv m n$ and $\hat{\omega} \equiv m\omega$ are dimensionless ``versions'' of the fundamental frequencies. In particular, since for dimensional reasons $\hat{\calM}(\hat{n},\hat{\omega},\nu)$ cannot depend on the total mass $m$, Eq.~\eqref{za_dcalMdma} can be recast in the form
\beq\label{z_calM}
	\av{z_a} = 1 + \frac{1}{2} \left( 1 \pm \Delta - 4 \nu \right) \biggl( \hat{\calM} + \nu \, \frac{\partial \hat{\calM}}{\partial \nu} \biggr) + \nu \, \biggl( \hat{\calM} + \hat{\omega} \, \frac{\partial \hat{\calM}}{\partial \hat{\omega}} + \hat{n} \, \frac{\partial \hat{\calM}}{\partial \hat{n}} \biggr) \, ,
\eeq
where the plus sign (respectively, minus sign) stands for particle 1 (respectively, particle 2), and we recall that $\Delta = (m_2 - m_1) / m = (1-4\nu)^{1/2}$ is the reduced mass difference.

We now restrict to the extreme mass-ratio limit $q \equiv m_1 / m_2 \ll 1$, and denote by $\av{z} \equiv \av{z_1}$ the average redshift of the small body. Although it is common, in the context of perturbation theory, to use the small mass ratio $q$ as an expansion parameter, Refs.~\cite{De.79,Sm.79,FiDe.84,Fa.al.04,Le.al.11,Sp.al2.11,Le.al2.12,Le.al.13,Na.13,Le2.14} showed that it is more advantageous to use the \textit{symmetric} mass ratio $\nu = q /(1+q)^2$ instead.\footnote{Note that this parameter appears naturally in the PN expansions of quantities that are symmetric under the exchange $1 \leftrightarrow 2$ of the bodies' labels; see \textit{e.g.} Eqs.~\eqref{R} and \eqref{n_omega}.} (Formally, $q = \nu + \calO(\nu^2)$.) Hence, to leading order beyond the test-particle approximation, we consider the expansions
\begin{subequations}\label{exp}
	\begin{align}
		\hat{E} &= E_{(0)} + \nu \, E_{(1)} + \calO(\nu^2) \, , \label{Eexp} \\
		\hat{L} &= L_{(0)} + \nu \, L_{(1)} + \calO(\nu^2) \, , \label{Lexp} \\
		\hat{R} &= R_{(0)} + \nu \, R_{(1)} + \calO(\nu^2) \, , \label{Rexp} \\
		\hat{\calM} &= \calM_{(0)} + \nu \, \calM_{(1)} + \calO(\nu^2) \, , \\
		\av{z} &= \av{z}_{(0)} + \nu \, \av{z}_{(1)} + \calO(\nu^2) \, ,
	\end{align}
\end{subequations}
where all the quantities are functions of the dimensionless frequencies $\hat{n}$ and $\hat{\omega}$. The variables with a subscript $(0)$ correspond to the geodesic values, while those with a subscript $(1)$ denote the conservative GSF corrections (at fixed frequencies).\footnote{In particular, $1 + E_{(0)}$ and $L_{(0)}$ coincide with the usual conserved (specific) energy and angular momentum associated with the stationarity and axisymmetry of the background Schwarzschild geometry.}

At zeroth order in the (symmetric) mass ratio $\nu$, Eq.~\eqref{z_calM} implies $\av{z}_{(0)} = 1 + \calM_{(0)}$, such that Eqs.~\eqref{M_calM}--\eqref{R_calM} yield the following relations between the leading-order contributions to the average redshift, binding energy, angular momentum and radial action integral:
\begin{subequations}\label{geo}
	\begin{align}
		E_{(0)} &= \av{z}_{(0)} - \hat{\omega} \, \frac{\partial \av{z}_{(0)}}{\partial \hat{\omega}} - \hat{n} \, \frac{\partial \av{z}_{(0)}}{\partial \hat{n}} - 1 \, , \label{E0} \\
		L_{(0)} &= - \frac{\partial \av{z}_{(0)}}{\partial \hat{\omega}} \, , \quad R_{(0)} = - \frac{\partial \av{z}_{(0)}}{\partial \hat{n}} \, .
\end{align}
\end{subequations}
Adjusting notations, these formulas agree with Eqs.~(A2)--(A3) and (A7) of Ref.~\cite{Ak.al.15}. At the next-to-leading order, \eqref{z_calM} implies $\av{z}_{(1)} = 2 \calM_{(1)} - 2 \calM_{(0)} + \hat{\omega} \, (\partial \calM_{(0)} / \partial \hat{\omega}) + \hat{n} \, (\partial \calM_{(0)} / \partial \hat{n})$, from which we deduce that the GSF contributions to the binding energy, angular momentum and radial action integral are given in terms of the GSF contribution to the average redshift (as well as geodesic quantities) by
\begin{subequations}\label{GSFs}
	\begin{align}
		E_{(1)} &= \frac{1}{2} \left( \av{z}_{(1)} - \hat{\omega} \, \frac{\partial \av{z}_{(1)}}{\partial \hat{\omega}} - \hat{n} \, \frac{\partial \av{z}_{(1)}}{\partial \hat{n}} + 2 E_{(0)} \right. \nonumber \\ &\left. \qquad\qquad\!\! + \; \hat{\omega}^2 \, \frac{\partial^2 \av{z}_{(0)}}{\partial \hat{\omega}^2} + 2 \hat{\omega} \hat{n} \, \frac{\partial^2 \av{z}_{(0)}}{\partial \hat{\omega} \partial \hat{n}} + \hat{n}^2 \, \frac{\partial^2 \av{z}_{(0)}}{\partial \hat{n}^2} \right) , \label{E1} \\
		L_{(1)} &= - \frac{1}{2} \left( \frac{\partial \av{z}_{(1)}}{\partial \hat{\omega}} + \frac{\partial \av{z}_{(0)}}{\partial \hat{\omega}} - \hat{\omega} \, \frac{\partial^2 \av{z}_{(0)}}{\partial \hat{\omega}^2} - \hat{n} \, \frac{\partial^2 \av{z}_{(0)}}{\partial \hat{n} \partial \hat{\omega}} \right) , \label{L1} \\
		R_{(1)} &= - \frac{1}{2} \left( \frac{\partial \av{z}_{(1)}}{\partial \hat{n}} + \frac{\partial \av{z}_{(0)}}{\partial \hat{n}} - \hat{n} \, \frac{\partial^2 \av{z}_{(0)}}{\partial \hat{n}^2} - \hat{\omega} \, \frac{\partial^2 \av{z}_{(0)}}{\partial \hat{\omega} \partial \hat{n}} \right) .
\end{align}
\end{subequations}
These relationships generalize Eqs.~(4) of Ref.~\cite{Le.al2.12} to generic bound (eccentric) orbits. Hence, if the function $\av{z}_{(1)}(\hat{\omega},\hat{n})$ is known, the leading-order corrections $E_{(1)}(\hat{\omega},\hat{n})$, $L_{(1)}(\hat{\omega},\hat{n})$ and $R_{(1)}(\hat{\omega},\hat{n})$ to the geodesic values can be computed for \textit{any} bound orbit. Such GSF data for $\av{z}_{(1)}$ is already available for a range of orbits with eccentricities $0 \lesssim e \lesssim 0.4$ \cite{BaSa.11,Ak.al.15,vMeSh.15}.

The relationships \eqref{GSFs} could easily be generalized up to second (or higher) order, yielding the expressions for the $\calO(\nu^2)$ contributions to the binding energy, angular momentum and radial action integral in \eqref{exp}, say $E_{(2)}(\hat{n},\hat{\omega})$, $L_{(2)}(\hat{n},\hat{\omega})$ and $R_{(2)}(\hat{n},\hat{\omega})$, in terms of the second order GSF contribution to the average redshift, say $\av{z}_{(2)}(\hat{n},\hat{\omega})$, and its frequency derivatives (as well as geodesic and first-order GSF quantities). Thanks to recent progress in formulating the second-order GSF \cite{Ro.06,De.12,Gr.12,Po.12,Po2.12,PoMi.14,Po.15}, numerical results for the second-order contribution $z_{(2)}(\hat{\omega})$ to the redshift for \textit{circular} orbits should soon become available \cite{Po.14}. Combining these data with the first law of binary mechanics would then yield the fully relativistic expressions for the functions $E_{(2)}(\hat{\omega})$ and $L_{(2)}(\hat{\omega})$, to be added to the geodesic and first-order contributions that were obtained in Ref.~\cite{Le.al2.12}.

\subsection{Schwarzschild separatrix and singular curve}\label{subsec:Schw}

Interestingly, Eqs.~\eqref{geo} and \eqref{GSFs} could be used to explore some conservative GSF effects on the motion of a self-gravitating particle on a bound orbit around a nonspinning black hole. For instance, one could compute the GSF-induced shift in the location of the Schwarzschild \textit{separatrix}, the curve that separates---in the relevant parameter space---between stable bound (eccentric) orbits and unstable (plunging) ones \cite{Cu.al.94}. (Along the separatrix, the radial period $P$ and the angular advance per orbit $\Phi$ both diverge, while their ratio $\Phi / P$ remains finite; therefore these orbits are characterized by a vanishing radial frequency $n$ at a fixed $\omega$.) Since all of the orbits that lie along the separatrix are marginally unstable, this calculation would generalize to eccentric orbits that of the shift of the Schwarzschild ISCO frequency induced by the conservative GSF \cite{BaSa.09,BaSa.10,Le.al2.12,Ak.al.12,Is.al.14}. Especially interesting would be the calculation of the GSF-induced shift in the frequency of the Schwarzschild innermost bound stable orbit (IBSO). Indeed, the ISCO and the IBSO are the end points of the separatrix.

However we note that, in order to compute the binding energy and angular momentum of such marginally unstable orbits, it may not be necessary to rely on eccentric-orbit GSF data for $\av{z}_{(1)}(\hat{\omega},\hat{n})$, in the limit where $\hat{n} \to 0$ at fixed $\hat{\omega}$. Indeed, in the test-mass approximation, it is known that to each orbit along the separatrix corresponds an \textit{unstable circular} timelike geodesic (homoclinic orbit \cite{LePe.09}) that shares the same binding energy and angular momentum. Assuming that this property still holds while taking into account the effect of self-interaction, one could simply combine the (circular-orbit restriction of the) first law of binary mechanics with the circular-orbit GSF data for $z_{(1)}(\hat{\omega})$ that is provided in Table IX of Ref.~\cite{Ak.al.12}.

Another key property of bound timelike geodesic orbits around a Schwarzschild black hole is the existence of a \textit{singular curve} in the parameter space, along which the Jacobian matrix of the transformation $(E_{(0)},L_{(0)}) \leftrightarrow (\omega,n)$ becomes singular \cite{BaSa.11}. This strong-field feature is closely related to the existence of a separatrix, and is responsible for the recently discovered phenomenon of \textit{isofrequency pairing} \cite{Wa.al.13}, namely, the existence of physically distinct orbits (having different $E_{(0)}$ and $L_{(0)}$) that share the same frequencies $\omega$ and $n$. Going beyond the test-particle approximation, we find using Eqs.~\eqref{exp}--\eqref{GSFs} that the Jacobian determinant of the transformation $(\hat{E},\hat{L}) \leftrightarrow (\hat{\omega},\hat{n})$ is given by
\begin{align}\label{J}
	J \equiv \biggl| \frac{\partial (\hat{E},\hat{L})}{\partial (\hat{\omega},\hat{n})} &\biggr| = \hat{n} \left\{ \biggl( \frac{\partial^2 \av{z}_{(0)}}{\partial \hat{\omega} \partial \hat{n}} \biggr)^2 -  \frac{\partial^2 \av{z}_{(0)}}{\partial \hat{\omega}^2} \frac{\partial^2 \av{z}_{(0)}}{\partial \hat{n}^2} \right\} \nonumber \\ &\!\! + \nu \, \hat{n} \, \Biggl\{ \frac{\partial^2 \av{z}_{(0)}}{\partial \hat{\omega} \partial \hat{n}} \left( \frac{\partial^2 \av{z}_{(1)}}{\partial \hat{\omega} \partial \hat{n}} - \hat{\omega} \, \frac{\partial^3 \av{z}_{(0)}}{\partial \hat{\omega}^2 \partial \hat{n}} - \hat{n} \, \frac{\partial^3 \av{z}_{(0)}}{\partial \hat{\omega} \partial \hat{n}^2} \right) \nonumber \\ &\qquad\; - \frac{1}{2} \frac{\partial^2 \av{z}_{(0)}}{\partial \hat{\omega}^2} \left( \frac{\partial^2 \av{z}_{(1)}}{\partial \hat{n}^2} - \hat{\omega} \, \frac{\partial^3 \av{z}_{(0)}}{\partial \hat{\omega} \partial \hat{n}^2} - \hat{n} \, \frac{\partial^3 \av{z}_{(0)}}{\partial \hat{n}^3} \right) \nonumber \\ &\qquad\; - \frac{1}{2} \frac{\partial^2 \av{z}_{(0)}}{\partial \hat{n}^2} \left( \frac{\partial^2 \av{z}_{(1)}}{\partial \hat{\omega}^2} - \hat{\omega} \, \frac{\partial^3 \av{z}_{(0)}}{\partial \hat{\omega}^3} - \hat{n} \, \frac{\partial^3 \av{z}_{(0)}}{\partial \hat{\omega}^2 \partial \hat{n}} \right) \Biggr\} + \calO(\nu^2) \, .
\end{align}
In the test-mass limit $\nu \to 0$, the Schwarzschild singular curve $\hat{\omega} = \hat{\omega}_s(\hat{n})$ along which $J \to \infty$ can be computed from the condition $\bigl[ (\partial^2 \av{z}_{(0)} / \partial \hat{\omega} \partial \hat{n})^2 - (\partial^2 \av{z}_{(0)} / \partial \hat{\omega}^2) (\partial^2 \av{z}_{(0)} / \partial \hat{n}^2) \bigr]^{-1} \!\! = 0$. Moreover, once the second-order partial derivatives of $\av{z}_{(1)}(\hat{n},\hat{\omega})$ evaluated along the curve $\hat{\omega} = \hat{\omega}_s(\hat{n})$ are known, it will become straightforward, thanks to Eq.~\eqref{J}, to compute the GSF-induced shift in the location of the Schwarzschild singular curve.

Carrying out the applications outlined above requires obtaining GSF data for $\av{z}_{(1)}(\hat{n},\hat{\omega})$ (as well as its partial derivatives) in a region of the parameter space that remains challenging for state-of-the-art GSF codes, namely for strong-field geodesic orbits that lie deep in the ``zoom-whirl'' \cite{GlKe.02} regime.\footnote{Unbound orbits close to the separatrix also play an important role in a scenario that explores the possibility of overspinning a nearly extremal Kerr black hole by means of a particle plunging from infinity \cite{CoBa.15}.} Nevertheless recent work suggests that such computations could become tractable in the foreseeable future, for instance by using a method relying on Green functions and worldline integrations \cite{Wa.al.14}.

\subsection{Noncircular conservative EOB dynamics}\label{subsec:EOB}

As pointed out in Ref.~\cite{Da.10}, GSF results can be used to inform the EOB model, including for strong-field orbits, by providing the exact expressions for the $\calO(\nu)$ contributions to the metric potentials that enter the conservative piece of the EOB dynamics. By construction, the conservative part of the EOB dynamics derives from the Hamiltonian \cite{BuDa.99}
\beq\label{H_EOB}
	H_\text{EOB} = m \, \sqrt{1 + 2 \nu \, \bigl( \hat{H}_\text{eff} - 1 \bigr)} \, ,
\eeq
where $m = m_1 + m_2$ is the total mass of the binary, while $\hat{H}_\text{eff} = H_\text{eff} / \mu$ denotes the ``effective Hamiltonian'' of an ``effective particle'' of mass $\mu = m_1 m_2 / m$ that follows a timelike geodesic (modulo quartic and higher-order terms in the momentum) in the ``effective metric''
\beq
	\ud s^2_\text{eff} = g_{\alpha\beta}^\text{eff}(x) \, \ud x^\alpha \ud x^\beta = - A(r;\nu) \, \ud t^2 + B(r;\nu) \, \ud r^2 + r^2 \left( \ud \theta^2 + \sin^2{\theta} \, \ud \varphi^2 \right) .
\eeq
This effective metric is a deformation of the Schwarzschild geometry of a black hole of mass $m$, with deformation parameter $\nu = \mu / m$, such that $A = B^{-1} = 1 - 2m/r$ in the test-mass limit $\nu \to 0$.\! The effective Hamiltonian entering the EOB Hamiltonian \eqref{H_EOB} reads \cite{Da.al3.00,Da.al.15}
\beq\label{H_eff}
	H_\text{eff}(r,p_r,L_0) = \sqrt{A(r;\nu) \left( \mu^2 + \frac{L_0^2}{r^2} + \frac{p_r^2}{B(r;\nu)} + Q(r,p_r;\nu) \right)} \, ,
\eeq
where $L_0$ is the conserved EOB angular momentum and $Q(r,p_r;\nu)$ is a function that controls the deviation from geodesic motion in the effective metric $g_{\alpha\beta}^\text{eff}(x)$. Up to 3PN order, we have $\hat{Q} = (8 - 6\nu) \, \nu \, \hat{p}_r^4 / \hat{r}^2$ where, for convenience, we introduced the rescaled variables $\hat{r} \equiv r / m$, $\hat{p}_r \equiv p_r / \mu$ and $\hat{Q} \equiv Q / \mu^2$. The 4PN contribution to the EOB potential $\hat{Q}$, and those entering the expressions for $A$ and $B$, have recently been computed in Ref.~\cite{Da.al.15}. Following Ref.~\cite{Da.10}, we shall assume that the fact that $\hat{Q}$ depends solely on $\hat{p}_r$ and $\hat{r}$, and vanishes (at least) like $\hat{p}_r^4$ when $\hat{p}_r \to 0$, remains true at higher PN orders; see also \cite{Ba.al.12}. Moreover, we shall restrict our analysis to \textit{mildly eccentric} orbits, for which the contribution $\calO(\hat{p}_r^6)$ (and higher orders) to $\hat{Q}$ can be neglected.

Now, in the extreme mass-ratio limit $\nu \ll 1$, the potentials $A(u;\nu)$, $B(u;\nu)$ and $\hat{Q}(u,\hat{p}_r;\nu)$ that enter the conservative EOB dynamics can be expanded as
\begin{subequations}\label{A-D-Q}
	\begin{align}
		A &= 1 - 2u + \nu \, a(u) + \calO(\nu^2) \, , \\
		\bar{D} &= 1 + \nu \, \bar{d}(u) + \calO(\nu^2) \, , \\
		\hat{Q} &= \nu \, q(u) \, \hat{p}_r^4 + \calO(\nu^2) \, ,
	\end{align}
\end{subequations}
where $u \equiv 1 / \hat{r}$ is a dimensionless measure of the gravitational potential, and $\bar{D} \equiv {(AB)}^{-1}$. For circular orbits, all of the information about the conservative EOB dynamics is encoded in the metric coefficient $g_{tt}^\text{eff} = - A$. By combining the first law of Paper I with GSF data for the redshift variable, Refs.~\cite{Le.al2.12,Ba.al.12} could compute the function $a(u)$ for all $0 < u \leq 1/5$. Making use of additional GSF data for circular orbits, all the way down to the Schwarzschild light-ring, Akcay \textit{et al.} \cite{Ak.al.12} then completed the determination of the exact linear-in-$\nu$ contribution to the EOB potential $A(u;\nu)$ for all $0 < u < 1/3$. Moreover, Damour \cite{Da.10} showed that, for slightly noncircular orbits, the GSF contribution to the invariant relationship $K(\omega)$ is related to a linear combination of the functions $a(u)$ and $\bar{d}(u)$. This GSF effect was later computed numerically in Ref.~\cite{Ba.al.10}, which allowed to determine the linear-in-$\nu$ contribution to the EOB potential $\bar{D}(u;\nu)$ [and hence $B(u;\nu)$] for all $0 < u \leq 1/6$ \cite{Ba.al.12,Ak.al.12}. However, the function $q(u)$ that encodes the $\calO(\nu \, \hat{p}_r^4)$ contribution to the EOB potential $\hat{Q}(u,\hat{p}_r;\nu)$ has, so far, remained entirely unconstrained (except for the leading 3PN and subleading 4PN terms when $u \to 0$). We shall now show how, by combining Eqs.~\eqref{GSFs} with GSF data for the relation $\av{z}_{(1)}(\hat{n},\hat{\omega})$ (for mildly eccentric orbits), the function $q(u)$ can be determined.

Having neglected terms $\calO(\hat{p}_r^6)$ (and higher) in the EOB potential $\hat{Q}$, and working to linear order in $\nu$, the effective EOB Hamiltonian \eqref{H_eff} squared can be expressed as a quadratic polynomial in the radial momentum squared as
\beq
	\hat{H}^2_\text{eff}(u,\hat{p}_r,\hat{L}_0) = A(u;\nu) \left( 1 + u^2 \hat{L}_0^2 + A(u;\nu) \bar{D}(u;\nu) \, \hat{p}_r^2 + \nu \, q(u) \, \hat{p}_r^4 \right) .
\eeq
Inserting the expansions \eqref{A-D-Q}, and solving for the radial momentum yields $\hat{p}_r$ as a function of $u = 1 / \hat{r}$ and the conserved quantities $\hat{\calE}_0 \equiv \hat{H}_\text{eff}$ and $\hat{L}_0 \equiv L_0 / (m \mu)$. Splitting the resulting expression into $\calO(\nu^0)$ and $\calO(\nu)$ contributions, i.e., $\hat{p}_r = \hat{p}_r^{(0)} + \nu \, \hat{p}_r^{(1)} + \calO(\nu^2)$, we find
\begin{subequations}\label{pr2}
	\begin{align}
		\hat{p}_r^{(0)}(u;\hat{\calE}_0,\hat{L}_0) &= \frac{\sqrt{\hat{\calE}_0^2 - (1-2u)(1+u^2\hat{L}_0^2)}}{1-2u} \, , \label{pr2(0)} \\
		2 \hat{p}_r^{(1)}(u;\hat{\calE}_0,\hat{L}_0) &= - \frac{\hat{\calE}_0^2}{\hat{p}_r^{(0)}} \, \frac{a(u)}{{(1-2u)}^3} - \hat{p}_r^{(0)} \biggl( \bar{d}(u) + \frac{a(u)}{1-2u} + \bigl( \hat{p}_r^{(0)} \bigr)^2 \, \frac{q(u)}{1-2u} \biggr) \, . \label{pr2(1)}
	\end{align}
\end{subequations}
As expected, Eq.~\eqref{pr2(0)} coincides with the well-known expression for the radial momentum of a test particle of energy $\hat{\calE}_0$ and angular momentum $\hat{L}_0$ in orbit around a nonspinning black hole of mass $m$. The correction term \eqref{pr2(1)} involves, additionally, the known functions $a(u)$ and $\bar{d}(u)$, as well as the unknown function $q(u)$ that we intend to constrain.

For bound orbits, the turning points of the radial motion correspond to the two smallest real, postive and finite roots of the equation $\hat{p}_r(u;\hat{\calE}_0,\hat{L}_0) = 0$, say $u_- = 1 / \hat{r}_+$ and $u_+ = 1 / \hat{r}_-$. To linear order in $\nu$, these can be written as
\beq\label{u+-}
	u_\pm = u_\pm^{(0)} + \nu \, u_\pm^{(1)} + \calO(\nu^2) \, ,
\eeq
where the geodesic values $u_\pm^{(0)}$ are known, for any given $\hat{\calE}_0,\hat{L}_0$, by solving $\hat{p}_r^{(0)}(u_\pm^{(0)};\hat{\calE}_0,\hat{L}_0) = 0$. Expanding $\hat{p}_r(u;\hat{\calE}_0,\hat{L}_0) = 0$ to linear order in $\nu$, the mass-ratio corrections $\nu \, u_\pm^{(1)}$ are found to be solutions of $u_\pm^{(1)} (\partial \hat{p}_r^{(0)} \! / \partial u) (u_\pm^{(0)}) + \hat{p}_r^{(1)}(u_\pm^{(0)}) = 0$, the expression of which requires the knowledge of the unknown values for $q(u_\pm^{(0)})$. \!However, as we shall see shortly the knowledge of $u_\pm^{(1)}$ will not be necessary to constrain the function $q(u)$.

For any given $\hat{\cal{E}}_0$ and $\hat{L}_0$, we may now integrate the radial momentum \eqref{pr2} over a radial period to obtain the reduced EOB radial action (recall that $u = 1 / \hat{r}$)
\beq\label{R_EOB}
	\hat{R}_\text{EOB} \equiv \frac{1}{2\pi} \oint \hat{p}_r \, \ud \hat{r} = \frac{2}{2\pi} \int_{u_-}^{u_+} \! \Bigl( \hat{p}_r^{(0)}(u;\hat{\cal{E}}_0,\hat{L}_0) + \nu \, \hat{p}_r^{(1)}(u;\hat{\calE}_0,\hat{L}_0) \Bigr) \, \frac{\ud u}{u^2} + \calO(\nu^2) \, .
\eeq
Using the formula \eqref{u+-}, this integral splits into three contributions, the ranges of integration of which vary from $u_-^{(0)} + \nu \, u_-^{(1)}$ to $u_-^{(0)}$, from $u_-^{(0)}$ to $u_+^{(0)}$, and from $u_+^{(0)}$ to $u_+^{(0)} + \nu \, u_+^{(1)}$. Expanding to linear order in $\nu$, the first and last integrals will contribute terms proportional to $\nu \, u_-^{(1)} \hat{p}_r^{(0)}(u_-^{(0)})$ and $\nu \, u_+^{(1)} \hat{p}_r^{(0)}(u_+^{(0)})$, respectively. But these vanish by virtue of the definition of the radial turning points in the test-mass limit, such that the EOB radial action \eqref{R_EOB} can be computed by integrating over the \textit{geodesic} interval between $u_-^{(0)}(\hat{\cal{E}}_0,\hat{L}_0)$ and $u_+^{(0)}(\hat{\cal{E}}_0,\hat{L}_0)$. Hence, for any given $\hat{\cal{E}}_0$ and $\hat{L}_0$, the reduced EOB radial action may be written in the form $\hat{R}_\text{EOB} = R_{(0)} + \nu \, \Delta R + \calO(\nu^2)$, where $R_{(0)}$ is the radial action for a test particle orbiting a Schwarzschild black hole of mass $m$, and
\beq\label{DeltaR}
	\Delta R \equiv \frac{1}{\pi} \int_{u_-^{(0)}}^{u_+^{(0)}} \! \hat{p}_r^{(1)}(u;\hat{\cal{E}}_0,\hat{L}_0) \, \frac{\ud u}{u^2} \, .
\eeq

Next, recall that the EOB model is (by construction) built upon the identification of the ``on-shell'' Hamiltonian \eqref{H_EOB} with the total mass-energy of the binary, $H_\text{EOB} = M$, as well as the identification of the EOB angular momentum entering the effective EOB Hamiltonian \eqref{H_eff} with that of the real two-body system, i.e. $L_0 = L$ \cite{BuDa.99}. Therefore, in terms of reduced variables, we find by inverting \eqref{H_EOB} with $M = m + \mu \hat{E}$ that the EOB and real radial actions must obey
\beq\label{R_EOB-R}
	\hat{R}_\text{EOB} (\hat{\calE}_0,\hat{L}_0) = \hat{R}(\hat{E},\hat{L}) \, , \quad \text{where} \quad
	\begin{cases}
		\hat{\calE}_0 = 1 + \hat{E} + (\nu/2) \, \hat{E}^2 \, , \\
		\hat{L}_0 = \hat{L} \, .
	\end{cases}
\eeq
The perturbative equalities $\hat{R} = \hat{R}_\text{EOB} + (\nu/2) \hat{E}^2 (\partial \hat{R} / \partial \hat{E}) + \calO(\nu^2)$ and $\hat{R}_\text{EOB} = R_{(0)} + \nu \, \Delta R + \calO(\nu^2)$ can be combined, together with the relationship $\partial \hat{R} / \partial \hat{E} = \hat{n}^{-1}$ and $\partial \hat{R} / \partial \hat{L} = - \Phi / 2\pi$, as well as Eqs.~\eqref{Eexp} and \eqref{Lexp} to expand $\hat{R}(\hat{E},\hat{L})$ at fixed frequencies $\hat{n}$ and $\hat{\omega}$, to obtain
\beq\label{Rhat}
	\hat{R} = R_{(0)} + \nu \left\{ \frac{1}{\hat{n}} \bigl( E_{(1)} - \hat{\omega} L_{(1)} \bigr) + \frac{E_{(0)}^2}{2\hat{n}} + \Delta R \right\} + \calO(\nu^2) \, .
\eeq
Combining Eqs.~\eqref{E0} and \eqref{GSFs} yields $E_{(1)} - \hat{\omega} L_{(1)} = \hat{n} R_{(1)} + \frac{1}{2} \bigl( \av{z}_{(1)} + \av{z}_{(0)} + E_{(0)} - 1 \bigr)$. Hence, using Eq.~\eqref{Rexp} we note that the sum of the first two terms in the right-hand side of Eq.~\eqref{Rhat} cancels out the left-hand side. Since the $\calO(\nu)$ contribution must vanish identically, we obtain the key identity
\beq\label{identity}
	2 \hat{n} \, \Delta R + \av{z}_{(1)} + \av{z}_{(0)} + E_{(0)} + E_{(0)}^2 = 1 \, .
\eeq
In summary, we have shown that---for any given $\hat{n}$ and $\hat{\omega}$---the definite radial integral \eqref{DeltaR}, which involves the functions $a(u)$, $\bar{d}(u)$ and $q(u)$, is equal to a known function of the fundamental frequencies (assuming that the GSF contribution $\av{z}_{(1)}(\hat{n},\hat{\omega})$ is known). In particular, Eq.~\eqref{identity} shows that, for mildly eccentric orbits, all of the information about the linear-in-$\nu$ contributions to the EOB potentials is encoded in the function $\av{z}_{(1)}(\hat{n},\hat{\omega})$. Using Eqs.~\eqref{pr2} and \eqref{identity}, we can now easily deduce explicit expressions for $a(u)$, $\bar{d}(u)$ and $q(u)$.

To do so, it is quite convenient to parameterize the orbit in terms of a ``semi-latus rectum'' $p$ and an ``eccentricity'' $e$, instead of the frequencies $\hat{n}$ and $\hat{\omega}$, and to perform an expansion in the limit $e \to 0$, which is well adapted to the mildly eccentric orbits that we are considering. Following Darwin \cite{Da.61}, we parameterize the radial motion using the ``relativistic anomaly'' $\chi$ via
\beq\label{chi}
	\hat{r}(\chi) = \frac{p}{1 + e \cos{\chi}} \quad \Longleftrightarrow \quad u(\chi) = v \left( 1 + e \cos{\chi} \right) ,
\eeq
where $v \equiv 1/p$ is the inverse semi-latus rectum and $0 \leq e < 1$ the eccentricity. In terms of the parameters $(v,e)$, the boundaries of the integral \eqref{DeltaR} simply read $u_\pm^{(0)} = v \, (1 \pm e)$. Adjusting notations, the expressions for $\hat{n}(v,e)$, $\hat{\omega}(v,e)$, $\av{z}_{(0)}(v,e)$, $E_{(0)}(v,e)$ and $L_{(0)}(v,e)$ are given, \textit{e.g.}, in Eqs.~(2.4)--(2.10) of Ref.~\cite{Ak.al.15}. All of these relationships can be computed analytically, as perturbative expansions in powers of $e$. Moreover, performing the change of variable $u \to \chi$ and substituting for \eqref{chi} in Eq.~\eqref{DeltaR} [with \eqref{pr2}], the integral $\Delta R(E_{(0)}(v,e),L_{(0)}(v,e))$ can also be computed analytically, in the small-eccentricity limit. The result, which we have obtained up to $\calO(e^4)$, is too cumbersome to be displayed here. Finally, in the small-$e$ limit, the GSF contribution to the generalized redshift can be expanded as
\beq
	\av{z}_{(1)}(v,e) = z_{(1)}(v) + \frac{e^2}{2!} \, \av{z}^{e^2}_{(1)}(v) + \frac{e^4}{4!} \, \av{z}^{e^4}_{(1)}(v) + o(e^4) \, ,
\eeq
where we used the notations $z_{(1)}(v) \equiv \lim_{e \to 0} \, \av{z}_{(1)}(v,e)$ , $\av{z}^{e^2}_{(1)}(v) \equiv \lim_{e \to 0} \partial^2 \av{z}_{(1)}(v,e) / \partial e^2$, and $\av{z}^{e^4}_{(1)}(v) \equiv \lim_{e \to 0} \partial^4 \av{z}_{(1)}(v,e) / \partial e^4$. The terms $\calO(e)$ and $\calO(e^3)$ must vanish, as can be checked explicitly from the absence of such contributions in the identity \eqref{identity}.

Equations \eqref{pr2} show that $\hat{p}_r^{(0)} \sim e$ and $\hat{p}_r^{(1)} \sim 1/e$ in the limit where $e \to 0$. Therefore, the leading-order contribution in Eq.~\eqref{DeltaR} is $\calO(e^0)$. Similarly, the variables $\hat{n}$, $\av{z}_{(0)}$, $\av{z}_{(1)}$ and $E_{(0)}$ all contribute at leading $\calO(e^0)$. At that order, the identity \eqref{identity} implies
\beq\label{a}
	a(v) = \sqrt{1-3v} \; z_{(1)}(v) - v \left( 1 + \frac{1-4v}{\sqrt{1-3v}} \right) .
\eeq
This formula agrees with the result obtained in \cite{Ba.al.12} (see (2.14) therein). While the derivation given in Ref.~\cite{Ba.al.12} required integrating an ordinary differential equation, our derivation yields in a straightforward manner the algebraic relation between $a$ and $z_{(1)}$.

At the next $\calO(e^2)$, the identity \eqref{identity} provides a relationship between the functions $\bar{d}(v)$, $a(v)$, $a'(v) \equiv \ud a / \ud v$, $a''(v) \equiv \ud^2 a / \ud v^2$ and $\av{z}^{e^2}_{(1)}(v)$. Substituting for Eq.~\eqref{a}, we find
\begin{align}\label{d}
	\bar{d}(v) &= \frac{v \left( 7 - \frac{141}{4} v + 45 v^2 \right)}{2 \, {(1-3v)}^{5/2}} - \frac{1 - \frac{21}{8} v}{{(1-3v)}^{3/2}} \, z_{(1)}(v) + \frac{2 - \frac{51}{2} v + 101 v^2 -132 v^3}{{(1-6v)}^2\sqrt{1-3v}} \, z'_{(1)}(v) \nonumber \\ &\qquad\quad - \frac{v \, (1-2v) \, \sqrt{1-3v}}{2 \, (1-6v)} \, z''_{(1)}(v) + \frac{(1-2v) \, \sqrt{1-3v}}{v \, (1-6v)} \, \av{z}^{e^2}_{(1)}(v) \, .
\end{align}
By making use of Eqs.~(4.49)--(4.50) and (B1) of Ref.~\cite{Ak.al.15}, one can determine the expressions for $z_{(1)}(v)$ and $\av{z}^{e^2}_{(1)}(v)$ in the weak-field limit $v \ll 1$, and then check that Eq.~\eqref{d} provides the correct 3PN expansion for the linear-in-$\nu$ contribution to the EOB potential $\bar{D}$, namely $\bar{d}(v) = 6 v^2 + 52 v^3 + \calO(v^4)$. Moreover, a comparison of the formula \eqref{d} with Eqs.~(5.21)--(5.25) of Ref.~\cite{Da.10} (together with \eqref{a} here) yields a relation between the GSF contribution to the ratio $W \equiv (\hat{n} / \hat{\omega})^2$ (denoted $\rho$ in Ref.~\cite{Da.10}), and the quantities $z_{(1)}$ and $\av{z}^{e^2}_{(1)}$. It would be interesting to use GSF data for $\rho$, $z_{(1)}$ and $\av{z}^{e^2}_{(1)}$ to check this prediction.

At the next $\calO(e^4)$, we obtain a relation between $q(v)$, $\bar{d}(v)$, $\bar{d}'(v)$, $\bar{d}''(v)$, $a(v)$, $a'(v)$, $a''(v)$, $a'''(v)$, $a''''(v)$ and $\av{z}^{e^4}_{(1)}(v)$. Substituting for Eqs.~\eqref{a} and \eqref{d}, we get
\begin{align}\label{q}
	q(v) &= \frac{9}{8} \frac{v \, {(1-2v)}^2}{{(1-3v)}^{7/2}} \left( 1 - \frac{47}{9} v + \frac{1349}{144} v^2 - \frac{71}{12} v^3 \right) - \frac{5}{16} \frac{v \, {(1-2v)}^2}{{(1-3v)}^{5/2}} \left( 1 - \frac{15}{8} v \right) z_{(1)}(v) \nonumber \\ &- \frac{7}{6} \frac{{(1-2v)}^3}{v \, {(1-6v)}^4\sqrt{1-3v}} \left( 1 - \frac{99}{4} v + \frac{3097}{14} v^2 - \frac{5214}{7} v^3 + 828 v^4 \right) \av{z}_{(1)}^{e^2}(v) \nonumber \\ &+ \frac{{(1-2v)}^4 {(1-3v)}^{3/2}}{9 v^2 {(1-6v)}^2} \, \av{z}_{(1)}^{e^4}(v) + \frac{2 {(1-2v)}^2}{v \, {(1-3v)}^{3/2} {(1-6v)}^5} \left( 1 - \frac{100}{3} v + \frac{22963}{48} v^2 \right. \nonumber \\ &\qquad \left. - \, \frac{372085}{96} v^3 + \frac{467057}{24} v^4 - \frac{185935}{3} v^5 + \frac{243789}{2} v^6 - \frac{269793}{2} v^7 + 64188 v^8 \right) z'_{(1)}(v) \nonumber \\ &+ \frac{{(1-2v)}^4 \sqrt{1-3v}}{v \, {(1-6v)}^3} \left( 1 - \frac{8}{3} v \right) \! \left( 1 - \frac{15}{2} v \right) {\av{z}_{(1)}^{e^2}}'(v) - \frac{{(1-2v)}^4 {(1-3v)}^{3/2}}{6 \, {(1-6v)}^2} \, {\av{z}_{(1)}^{e^2}}''(v) \nonumber \\ &- \frac{7}{12} \frac{v \, {(1-2v)}^3}{{(1-6v)}^4\sqrt{1-3v}} \left( 1 - \frac{285}{28} v - \frac{299}{14} v^2 + \frac{1851}{7} v^3 - \frac{2790}{7} v^4 \right) z''_{(1)}(v) \nonumber \\ & - \frac{v}{6} \, \sqrt{1-3v} \, \frac{{(1-2v)}^4}{{(1-6v)}^3} \left( 1 - \frac{25}{2} v + 24 v^2 \right) z'''_{(1)}(v) + \frac{v^2}{24} \frac{{(1-2v)}^4 {(1-3v)}^{3/2}}{{(1-6v)}^2} \, z''''_{(1)}(v) \, .
\end{align}
We checked that this expression reproduces the known 3PN result for the linear-in-$\nu$ contribution to the EOB potential $Q$, namely $q(v) = 8 v^2 + \calO(v^3)$. Using GSF data for $z_{(1)}$ and $\rho$, the authors of Refs.~\cite{Ba.al.10,Ak.al.12} devised accurate global fits for the functions $a(v)$ and $\bar{d}(v)$, that are based on simple analytic models. While $a(v)$ is known for all $0 < v < 1/3$, the function $\bar{d}(v)$ has only been determined over the range $0 < v \leq 1/6$. Using additional GSF data for $\av{z}^{e^2}_{(1)}$ and $\av{z}^{e^4}_{(1)}$, together with Eq.~\eqref{q}, one could similarly construct a global fit for $q(v)$.

Finally, we note that the knowledge of the GSF contributions to the invariant relationships $M(n,\omega)$ and $L(n,\omega)$ [from Eq.~\eqref{GSFs} and GSF data for $\av{z}_{(1)}(\hat{n},\hat{\omega})$] could be used to determine the exact, linear-in-$\nu$ contribution to the EOB potential $\hat{Q}(u,\hat{p}_r;\nu)$ for \textit{generic} bound orbits, and not merely the function $q(u)$ that encodes the $\calO(\hat{p}_r^4)$ contribution therein \cite{Da.10}. This task is left to future work.

\acknowledgments

It is a pleasure to thank L. Barack, T. Hinderer and J. Vines for useful discussions. This research was supported by a Marie Curie FP7 Integration Grant (PCIG13-GA-2013-630210).

\bibliography{}

\end{document}